# Security of the DNS Protocol

Implementation and Weaknesses Analyses of DNSSEC

Kaouthar Chetioui, Ghizlane Orhanou, Said El Hajji, Abdelmajid Lakbabi

Laboratoire Mathématiques, Informatique et Applications

Université Mohammed V – Agdal

Faculté des Sciences - Rabat

kaoutharchetioui@gmail.com, ghizlane.orhanou@gmail.com, elhajji@fsr.ac.ma, lakbabi@gmail.com

**ABSTRACT**

Today, Internet offers many critical applications. So, it becomes very crucial for Internet service providers to ensure traceability of operations and to secure data exchange. Since all these communications are based on the use of the Domain Name System (DNS) protocol, it becomes necessary to think to enhance and secure it by proposing a secure version of this protocol that can correct the whole or a part of the DNS protocol weaknesses and vulnerabilities. In this context, DNSsec was created by the IETF to ensure the integrity of DNS data and authentication of the source of such data. DNSsec is based on the key cryptography public to provide different security services. In the present paper, we will present first the DNS protocol and its weaknesses. After that, we will be interested in studying the DNSsec implementation and data exchange, and then give a deep analysis of its weaknesses.

## I. INTRODUCTION

The DNS (Domain Name System) is an indispensable part for the access to the Internet. The purpose of the DNS is the translation of domain names to IP addresses and vice versa; this service is called name resolution. However, this service has become very vulnerable to several types of attacks. So, it becomes necessary to look for a new method or new protocol that can remedy to this security problem. In this paper, we will see an overview of the name resolution via DNS and the vulnerabilities of this service. Then we will present the solutions proposed by the DNSSEC protocol, based especially on asymmetric cryptography, to secure DNS. In addition, we have performed the implementation and tested the DNSsec protocol. So we will present an example of our implementation in order to secure an original domain name. At the end, we will conclude with an analysis of the benefits and weaknesses of DNSSEC, and give some perspectives to improve the DNS security.

## II. PRESENTATION OF DNS

The DNS is a server that maps a domain name to an IP address. Initially, the Internet was formed by few machines. So name resolution was made easily from a file hosts which included all the areas names with their IP addresses. But now, given the large number of machines connected to the Internet, the resolution has become much more complex: for each area, there is a name server called "primary name server "and a "secondary name server" that replace the primary name server in case of malfunction or unavailability. The architecture of the DNS database is an inverted tree. The root is at the top represented by ".". Each node of the tree has a label that identifies it from its parent as shown in *"Fig. 1"*.

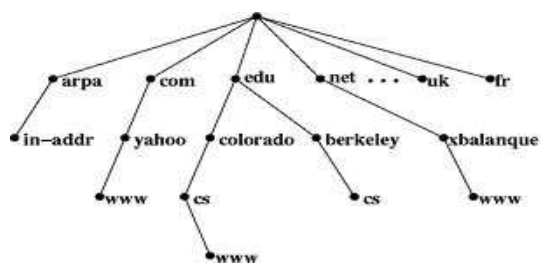

Figure 1. DNS Domain Name Hierarchy

Using the tree structure, it is possible to define areas of responsibility in the naming. Each one arrow connecting a node to a node can convey less information indicating who is responsible for the lower level [1]. In other words, the area behind the arrow, the father, contains information specifying which servers are with information about the son.

DNS has zone files that define it to be authoritative for some domains and slave for others and be configured to provide different operations for other domains or users. Zone files are numerous, the most important, in Linux system, are:

*a) / etc / named.conf: I*t contains the server configuration, and particularly the list of areas of which the server is authoritative (primary or secondary).

*b) /var/named/named.ca:* It contains the IP addresses of the Top level DNS servers that can inform a user from the root of the hierarchy.

When a client asks for the IP address of '*www.example.com*', the authoritative name server has to respond to queries. There are two types of queries defined for DNS systems:

- *Iterative mode*: In an iterative query, if the name server has the answer or if it is available in its cache, it will return it. If the name server doesn't have the answer, it will return any information to the next delegation level, that may be useful, but it will not make additional requests to other name server systems. All name servers must support iterative queries. *"Fig.2"* below illustrates an iterative query.

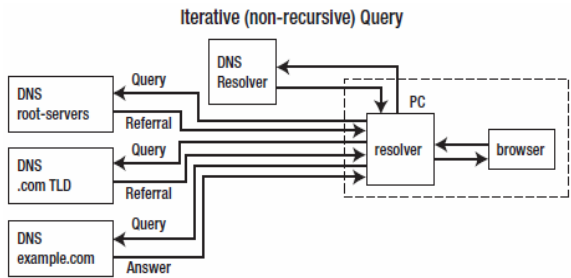

Figure 2. Iterative query [7]

- *Recursive mode*: Answering a query recursively may cause the name server to send multiple query transactions to a number of authoritative name servers in the DNS hierarchy in order to fully resolve the requested name. In the recursive query, the receiving name server will do all the necessary work to return the complete answer for the request. *"Fig.3"* illustrates the recursive query. We note that is not necessary for name servers to support recursive queries.

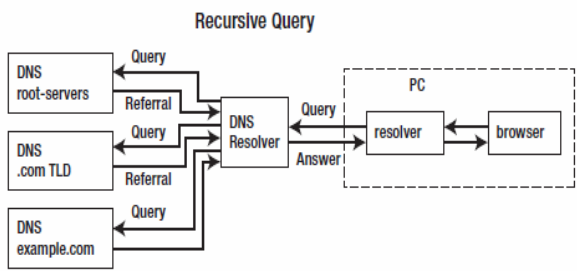

Figure 3. Recursive query [7]

III. WEAKNESSES IN DNS

As described above, the DNS allows machines to communicate over the Internet without knowing their IP addresses, but the original conception of this service does not include any service designed to secure transactions and the exchanged data. And therefore, the DNS remains highly vulnerable to attacks such as the creation or modification of messages, cache pollution, identity theft, etc [2]. *"Fig. 4"* below shows where attacks are possible and each possible vulnerability is marked with a question mark.

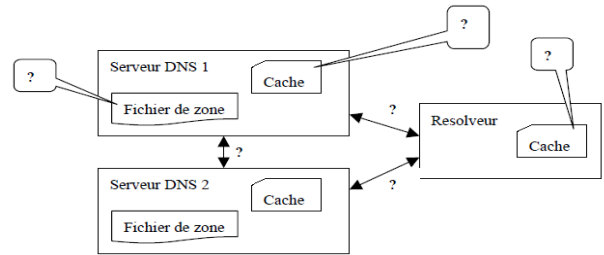

Figure 4. Weaknesses of DNS.

When there is a DNS resolution by a cache server, the attacks may be multiple. For example, if an attacker is on the same network as the legitimate client, he can return wrong answers by responding faster than the legitimate cache server.

A cache DNS server can return false answers or not to return answers at all. Recently, some providers have deliberately sent their users to advertising pages when the request domain doesn't exist [5].

In 2008, a new form of cache poisoning attack has occurred: the attack Kaminsky [5] is intended to poison a DNS cache server for an entire domain, not only a single registration as in the classic poisoning. It exploits two implementation problems of the DNS protocol: the number of expected requests, and the fact that the requests UDP source ports are always the same. For all these types of threads, a number of observers are suggesting that DNSSEC will become the default state for all zones.

We can see in *"Fig. 4"* that there are weaknesses located in DNS message exchanges, as well as potential intrusions in the zone file or the cache. There are four security services that may be provided by cryptography [7]:

*a) Authentification:* The data could only have come from a known source.

*b) Confidentiality:* Only the concerned parties of the communication can understand the messages sent between them.

*c) Integrity:* The data that is received by one party is the data that was sent by the other party.

*d) No-repudiation:* The sender of a message should not be able, subsequently deny having sent the message.

Authentication and data integrity are both absolutely necessary for DNS to deliver a safe service. Indeed, one must be able to ensure that the data received from a name server has not been altered and to verify the correct source of data**.** In the initial design of the DNS, no action has been taken to ensure the safety of the service and none of the security services listed above is provided.

## IV. PRESENTATION OF DNSSEC

As we have seen, the DNS has many faults and therefore it is very vulnerable to attack. Security problems of DNS concerns transaction security of DNS messages, data security, integrity, authentication and denial of services.

To remedy to this lack of security, the DNSSEC (Domain Name System Security Extensions) protocol is proposed. DNSSEC provides two essential security services to DNS: DNSSEC ensures data integrity and authentication of the source data. In order to implement them effectively, DNSSEC uses public key cryptography and digital signatures.

Two types of keys are identified for use in zone signing operations. The first type is called a *Zone Signing Key (ZSK)*, and the second type is called *Key Signing Key (KSK)*.

This extension uses new records (RR KEY, SIG, NSEC and DS) in addition to the original DNS records; it manages keys and signatures which are necessary for the use of a system of public key cryptography [3].

*a) DNSKEY RR recording*: The DNSKEY resource record stores all public key pairs. These public keys are needed to decrypt the encrypted information received as DNS replies.

*b) RRSIG record*: The RRSIG record results of the signing of RRsets made by the private key of the pair. Each domain name in a given area has at least one RRSIG record type.

*c) NSEC record*: is used by the DNSSEC protocol to check that if a negative response was received to a query host, it means that the target record doesn't exist; it's called proof of nonexistence and occasionally denial of existence [7].

*d) DS record*: It allows a parent zone to validate the KEY record of its child zone.

The resolution process by DNSSEC is as follows (*"Fig. 5"*):

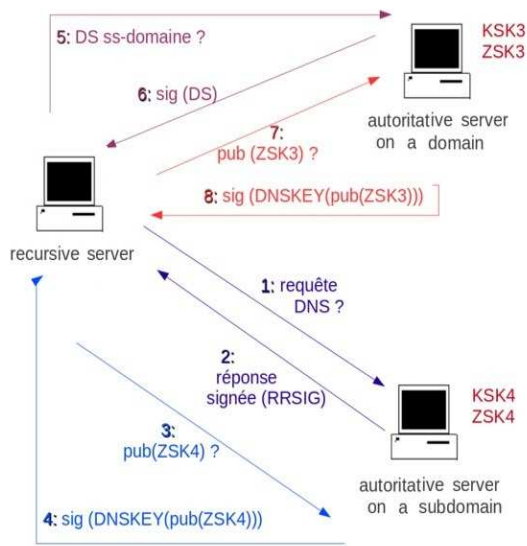

Figure 5.  Resolution process by DNSSEC [8].

Each record in the zone file is signed with the area private key. The signature generated is placed in a record type SIG. Each record type is related to the SIG record type which contains the signature.

If a third party changes the records in the zone file, the associated signature will not be correct, and the verification of this signature will indicates that the records were changed. This ensures the integrity of the records. In addition, only the server that has used the key can generate the correct signatures. If the signature verification is positive, it is possible to ensure that it is the owner of the private key that sent the records; the data source is as well authenticated.

The problem that remains also with DNS protocol is how to send a secure reply to a query on a domain name or records that don't exist. If a Resolver requests the address of a machine which doesn't exist, he will receive a response message containing an error code in the header of the DNS message and no signature is verifiable. To avoid this situation DNSSEC introduced the NSEC resource record.

This recording contains a bit vector indicating all types of existing name records associated with the NSEC record, and the next name is found in zone file according to the established order (lexicographic order on the labels: from the rightmost label to the leftmost label. An NSEC record also has its associated SIG record, so it is possible to infer securely that a name does not exist, if:

- It is between the domain name associated with the NSEC record and the name contained in the NSEC record,
- Record type does not exist,
- The bit associated with this type is 0 in the bit vector of the NSEC record received in response.

At this stage, all records that are necessary to the security of DNS exchanges seem to be defined. However, it was still possible for an attacker to generate a pair of keys (public / private) to sign false records. It will suffice then to intercept DNS requests and respond with false information. This introduces the problem of secure delegation of domain names. So, it's necessary to secure entry points in the chain of trust while maintaining the hierarchical structure of DNS.

For this, the RFC 4035 [4] proposes the following process: the zone administrator sends his daughter KEY record in the parent zone administrator, it signs the KEY record and returns the signature. The KEY record in the child zone is now signed by the key area of its parent zone. Thus, creating the signature of the KEY record in the child zone requires two messages. This procedure must be reproduced each time and a change is made on the registration of the child zone KEY or when the expiry date of the signature is exceeded.

Delegation Signer method uses a DS record stored in the parent zone, which contains among other things, the identifier of the public key of the child zone and a hash of the name of the child zone concatenated with the key value of the child zone. The DS record is constructed as illustrated in *"Fig. 6"* below.

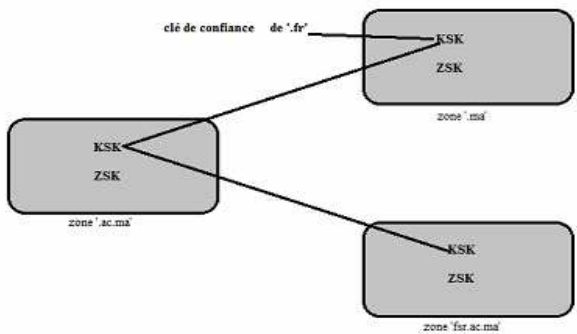

Figure 6. Chain of trust.

The administrator signs the child zone's Keyset with its private key, sends the KEY record containing the corresponding public key to the administrator of the parent zone. This creates the correspondent DS record that remains in the parent zone. Only one message is used to generate the record DS and a new message will be sent to change the DS record only when the key area of the child zone, which signed the KEY record, changes.

## V. IMPLEMENTATION OF DNSSEC

In the present section, we will present the different steps that we have followed to be able to implement and test the DNSSEC protocol.

As mentioned before, DNSSEC is the secure extension of DNS protocol, so its implementation should be made mandatory on a DNS server which already exists. So, first of all, we have started by installing the latest version of BIND 'version 9', then performed a standard configuration of the DNS server by creating a new domain name in the main file of DNS '/ etc / named.conf' (it is important to note that we are implementing DNSsec in Linux system) and zone files in the directory '/ var /named'. At the end of the configuration, we ensured that the server function properly by making resolution tests between the client and the server.

After having the satisfactory operation of the DNS, we can move on to the implementation of DNSSEC. This is complex task, so it needs to be done with care because any mistake can cause serious and costly failure in the network. To implement the DNSSEC, we have performed the following steps:

 *a)* *Create keys with dnssec-keygen*: This is the first step in the implementation of DNSSEC, which is to create ZSK and KSK that will serve us for the areas signature.

- *ZSK (Zone Signing Key):* is a key pair (private/public) used to sign the registers in the area.

- *KSK (Key Signing Key):* used to sign other keys and act as a 'Chain of Trust'.

*"Fig. 7"* below shows the commands used to generate the two keys ZSK and KSK.

```
#dnssec-keygen -a RSASHA1 -b 2000 -n ZONE domaine.ma
#dnssec-keygen -a RSASHA1 -b 2000 -n ZONE -f KSK domaine.ma
```

Figure 7. Example of ZSK and KSK generation.

This step requires the entropy to be accomplished and the choice of algorithms and keys size is left to administrators. In practice, 90% of areas are protected by RSA-SHA1 [5].

```
[root@CDC named]# cat Kdomaine.ma.+001+18235.key
; This is a zone-signing key, keyid 18235, for domaine.ma.
; Created: 20110711104640 (Mon Jul 11 10:46:40 2011)
; Publish: 20110711104640 (Mon Jul 11 10:46:40 2011)
; Activate: 20110711104640 (Mon Jul 11 10:46:40 2011)
domaine.ma.  IN DNSKEY  256 3 1 AwEAAbamzbwMtPRM3UG02zpv5UXSOxVDqhQuAKnDKGO
HlLTV6boLOcaE 7aUrvUcXROvUDU497QezS6hG4Vb5VJf0Icxbn8eMlNQ96kMgQFLgmoeK wg7xYZUm6
X18a/PHJEG0OpP+AZDq/+NmldgjZui6hWNKDAk+cms3/9Ha 3AEAzQy/xbtmcDV5Kgo84kaI94z/CQQr
xZ9d8Suhu1ZFjoS/ULhmE9T1 JmVQf0vc7a7u4sR0malwR7zVcXhh0JhwIfrHDklV58FIrBYB6FPKy3h
M K4H/dvWn94VOZBupQf9KQqEgzT1BRHAzT49/1BJPaPdkmEyf685jXeFH O2s=
```

Figure 8. Extract of the file 'kdomaine.ma' content (public)

```
Private-key-format: v1.3
Algorithm: 1 (RSA)
Modulus:
tqbNvAy09EzdQbTbOm/1RdI7FUOqFC4AqcMoY4eUtNXpugs5xoTtpSu9RxdE69QNTj3tB7N
LqEbhVv1U1/QhzFufx4yU1D3qQyBAUuCah4rCDvFh1SbpfXxr88ckQbQ6k/4BkOr/42aV2C
Nm6LqFY0oMCT5yazf/0drcAQDNDL/Fu22wNXkqA7ziRoj3jP8JBCvFn13xK6G7VkWOhL9Qu
GYT1PUmZVB/S9ztru7ixHSZqXBHvNVxeGHQmHAh+scOSVXnwUisFgHoU8rLEEwrgf929af3
hU5kG61B/0pCoSDNPUFEcDNPj3/UEk9o92SYTJ/rzmNd4Uc7aw==
PublicExponent: AQAB
PrivateExponent:
hgm26Q92I+ncXQG5+BJxcGGtFhfH0nNH7UzYcOHoUSTCFtNyHKskMpf8mRUMcPv2n7p2Hba
ICP+rEkZ6u+BehTef87LI7VBaf9RFIAzPmjRL4OdmsywRIROTrRwxET6P11cLEyxBiMGZqT
```

Figure 9. Extract of the file 'kdomaine.ma' content (private)

 *b)* *Inclusion of the keys in the zone files:* The second step is to add the previously created key in the zone file in order to be taken into account.

 *c)* *Signature of the zone with dnssec-signzone*: It is a very important step in configuring DNSSEC because during this steps, the areas are signed. To perform this step we need to type the command dnssec-signzone, as shown in *"Fig. 10"*. This command creates a new configuration file that has the same name as for the DNS, but with the extension '. signed' and containing all records in the area supplemented by the new DNSSEC recordings.

```
#dnssec-signzone -t -k KSK zonefile ZSK
```

Figure 10. Signature of the zone

```
[root@CDC named]# dnssec-signzone -t -k Kdomaine.ma.+001+45570
domaine.ma  Kdomaine.ma.+001+18235
Verifying the zone using the following algorithms: RSASHA1.
Zone signing complete:
Algorithm: RSAMD5: KSKs: 1 active, 0 stand-by, 0 revoked
                   ZSKs: 1 active, 0 stand-by, 0 revoked
domaine.signed
Signatures generated:                  23
Signatures retained:                    0
Signatures dropped:                     0
Signatures successfully verified:       0
Signatures unsuccessfully verified:     0
Runtime in seconds:                 0.141
Signatures per second:            162.838
```

Figure 11. Result of the zone signature

*d) Reconfiguration of 'named.conf'*: After the third step which create a new signed zone file, we must change the name of the old zone file in / etc / named.conf and replace it with the name of the new zone file with the extension '.signed', and in options, add 'dnssec-enableyes', as illustrated in *"Fig. 10"*.

```
file "domaine.ma.signed";
dnssec-enableyes;
```

Figure 12. Reconfiguration of 'named.conf'

*e) Restart BIND*: Restarting the service is an important step which needs to be done at the end of the configuration of the server to take account of the changes already made.

After having completed all the steps above, we have performed the necessary tests to ensure that the configuration was correctly done. This step was done initially at the server level and then at the client.

The command 'Dig' gives important information to ensure the good or the poor functioning of DNSSEC, so unlike the case with DNS protocol, the DNSSEC test with 'Dig' must provide the following results:

- Resource records used by the zone file.
- The flags 'qr aa rd ra' indicate that the signatures are validated and status 'NOERROR' also indicates that the configuration of DNSSEC is done without errors.

These results are illustrated in *"Fig. 10" below.*

```
[root@CDC ]    # dig domaine.ma +dnssec @localhost

; <<>> DiG 9.7.3 <<>>    domaine.ma +dnssec @localhost
;; global options: +cmd
;; Got answer:
;; ->>HEADER<<- opcode: QUERY, status: NOERROR, id: 42469
;; flags: qr aa rd ra; QUERY: 1, ANSWER: 2, AUTHORITY: 7, ADDITIONAL: 5

;; OPT PSEUDOSECTION:
; EDNS: version: 0, flags: do; udp: 4096
;; QUESTION SECTION:
; domaine.ma.              IN      A

;; ANSWER SECTION:
domaine.ma.        86400   IN      A       192.168.1.3
domaine.ma.        86400   IN      RRSIG   A 1 2 86400 20110812095331 20110
713095331 18235   domaine.ma.     WDOKIWPhATwSxIfOMQs7Chb57YQG+FLKrUr17HW9VAeeb
WyhUZwcXdrq nQTlz+PCWL/8nbF//QcgMju8YyHVLrECxTTRFQuiXJ8st3rpZZLyhPuj 8a2T0rGs3Uu
```

Figure 13. Test of DNSSEC with 'Dig'

In the other hand, once the DNSSEC server has functioned properly, we have performed the test of a DNSSEC client. But before the test, the client must contain the server key used by the DNSSEC server. To do this, we have executed the command:

'tail -n 1 KSK.key> / etc / trusted-key.key'

in the client to retrieve the key from the server. And then, we have performed the test again with the command 'Dig' and the test was conclusive.

In the DNSSEC test, there may be limits even if the configuration is done correctly; this problem can happen in the following cases:

- The used equipments don't support the file size of the area (at least ), in which case it will give a message that the maximum size is exceeded.
- For the clients using the operating system Microsoft Windows, the DNSSEC protocol is not supported apart from Microsoft Windows 7 which supports it.

## VI. WEAKNESSESS AND PROSPECTS OF DNSSEC

As we have discussed and presented above, the DNSSEC protocol have solved many security problems of the DNS protocol by providing authentication and data integrity but it is still vulnerable to some types of attacks. By adding resource records to secure transactions, the size of a DNSSEC zone file become seven times larger than that of a DNS file. Furthermore, the DNSSEC protocol will use TCP protocol and not UDP. This will cause an increase of the network load [2].

Moreover, up to now, there is no robust system that can face a Denial of Service attack. Therefore, given the size DNSSEC messages there is still vulnerable.

DNSSEC does not protect unsigned records, so we have to think about protecting the zone transfer by other techniques.

In addition, DNSSEC needs to be synchronized between the client and the entity making the synchronization, so during this phase, there could be an interception of files and the same problem can be arised during the keys renewal.

The deployment of DNSSEC on the internet is currently underway. The root zone is signed since July 2010. In March 2011, 65 top-level domains have a record DS in the root but they don't accept DS records for the stubs zone. Most registers don't add record DS but the most important ones should do it in 2011[5].

## VII. CONCLUSION

The DNSsec is a DNS security protocol that was adopted to enhance DNS security and ensure the authenticity and data integrity during data exchanges between DNS servers and between DNS clients and DNS servers. As we have presented in this paper, it uses asymmetric cryptography to provide maximum safety for the DNS servers. In addition, it uses new resource records and signatures to secure the zone transfer. We have been, also, interested in its implementation in a linux server, and the results of keys generation, zone signature and DNSsec operation have been given. At the end, and since DNSsec is still subject of many researches, we have carried out an analysis of the weaknesses and lacks of this security protocol and we found that, even if it ensures important security issues, it is still vulnerable to certain types of attacks.